\documentclass[journal]{IEEEtran}

\usepackage[T1]{fontenc}
\usepackage[latin9]{inputenc}
\usepackage{amsmath}
\usepackage{amssymb}
\usepackage{graphicx}
\usepackage{esint}

\usepackage{babel}

\begin{document}

\title{Density of Bloch states inside a one dimensional photonic crystal }

\author{Ebrahim Forati,~\IEEEmembership{Member,~IEEE}

\thanks{PO box 3043, Santa Clara, CA 95051, USA}
\thanks{email: forati@ieee.org}}

\maketitle
\begin{abstract}
The density of Bloch electromagnetic states inside a one dimensional
photonic crystal (1D PC) is formulated based on its dispersion relations.
The formulation applied to any anisotropic medium with known dispersion
relations and  iso-frequency surfaces. Using a practical
1D PC parameters in the visible range, the density of Bloch states for different modes 
are calculated.
\end{abstract}

\section*{Introduction}

The density of electromagnetic modes/states (DOS) is an important
quantity in statistical physics. It is often used as the degeneracy function for the energy levels
in thermal radiation studies such as Planck\textquoteright s blackbody
radiation \cite{kirchhoff1860relation,planck2013theory}. Consider the
distribution of  photons inside a large cavity in thermal equilibrium
with a solid matter at temperature T (a.k.a. cloud of photons). The
solid matter provides the mechanism to convert photons energies (i.e.
annihilate and create them) according to the temperature. Since photons
are bosons, the number of photons at each energy level ($\varepsilon_{i}=\hbar\omega$)
follows Bose-Einstein distribution as \cite{reichl1999modern}

\begin{equation}
n_{i}=\frac{g_{i}}{e^{-\alpha-\beta\varepsilon_{i}}-1}\label{eq:BE}
\end{equation}
where $g_{i}$ and $n_{i}$ are the degeneracy and photon number of
each energy level ($\varepsilon_{i}$), respectively. Equation (\ref{eq:BE})
is obtained by maximizing the number of micro-states in the system
(see \cite{reichl1999modern}). The constants $\alpha$ and $\beta$
are determined by enforcing the conditions $\sum n_{i}=N,$ and $\sum\varepsilon_{i}n_{i}=U$
where $N$ and $U$ are the total number and energy of bosons, respectively.
The later condition leads to $\beta=\frac{1}{k_{B}T}$ where $k_{B}$
is Boltzmann's constant, and the former does not apply to photons,
as they can be created and annihilated bythe solid
matter. This leads to setting $\alpha=0$. The modes degeneracy, $g_{i},$
is where the DOS is required. If there is no photons reflection at
the surface of the thermal radiator (i.e. an ideal black body,) the
DOS inside the thermal radiator is essentially the same as its ambient
medium. 

Alternatively, we could explain thermal emission by considering the
distribution of emitters inside a matter at temperature T, and relate
it to the radiational modes. An example of such methods is briefly
reviewed in the appendix, for self-consistency. 

If the emitting body is a 1D PC, the DOS consists of two contributions:
Bloch states which can propagate and depart a finite-size 1D PC, and the
wave-guided states which trap photons inside the 1D PC (mostly within higher permittivity
layers). A complete analysis of these states is performed in \cite{Forati_underreview}
based on Green's tensor approach. However, the density of Bloch states ($\mathrm{DOS_{Bloch}}$) can be calculated, exactly,  using their dispersion relations in the 1D PC.
This is the parameter which, depending on the geometry, can be used directly or indirectly as the degeneracy function in
thermal emission from 1D PCs. 

\section*{$\mathrm{DOS}_{\mathrm{Bloch}}\left(\omega\right)$
inside a 1D PC }

We start by formulating the DOS inside a medium with known dispersion
relations, and general non-spherical iso-frequency surfaces. The conventional
method of calculating the DOS is to consider a large rectangular cavity
with size L in all dimensions, filled with the medium, and with periodic
boundary conditions on its walls. Eigen solutions of the harmonic
electromagnetic wave equation, at frequency $\omega,$ in such geometry
are modes in the form 

\begin{equation}
F_{mnl}=C_{mnl}^{p}e^{-i\omega t}e^{i\left(\frac{2\pi m}{L}x+\frac{2\pi n}{L}y+\frac{2\pi l}{L}z\right)}
\end{equation}
where $F_{mnl}$ is an electric (magnetic) field component of the
mode and $C_{mnl}^{p}$ is its associated coefficient (which has both
frequency and geometry dependences.) The superscript $p$ in $C_{mnl}^{p}$
denotes the mode's polarization and can take two values (any two orthogonal
polarizations is correct, but transverse electric and transverse magnetic
modes along a cartesian coordinate axis are commonly chosen.) The
values in the triplet $\left(m,n,l\right)$ can be any positive integer
number subject to an equation obtained from the eigen value problem
\begin{equation}
f\left(\frac{2\pi m}{L},\frac{2\pi n}{L},\frac{2\pi l}{L},\omega\right)=0,\label{eq:disp1}
\end{equation}

Equation (\ref{eq:disp1}) is known as the dispersion equation after
defining and replacing the wave-vectors as 
\begin{equation}
\left(k_{x},k_{y},k_{z}\right)=\left(\frac{2\pi m}{L},\frac{2\pi n}{L},\frac{2\pi l}{L}\right).\label{eq:disp}
\end{equation}
In case of an isotropic homogenous material, equation (\ref{eq:disp1})
becomes $\left(\frac{2\pi m}{L}\right)^{2}+\left(\frac{2\pi n}{L}\right)^{2}+\left(\frac{2\pi l}{L}\right)^{2}=k_{0}^{2}\varepsilon_{r}\mu_{r}$.
The density of electromagnetic modes, $\mathrm{DOS}\left(\omega\right),$
is defined as the spectral density of the modes per cavity volume.
That is, $\mathrm{DOS}\left(\omega\right)d\omega$ is the number of
supported modes per volume in the interval $\left[\omega,\omega+d\omega\right]$,
\begin{equation}
\mathrm{DOS}\left(\omega\right)d\omega=\underset{\left\{ \left(m,n,l\right):f\left(\frac{2\pi m}{L},\frac{2\pi n}{L},\frac{2\pi l}{L},\omega<\omega^{\prime}\leq\omega+d\omega\right)=0\right\} }{\sum\frac{2}{L^{3}}}\label{eq:SD}
\end{equation}
where the factor of 2 accounts for the two possible orthogonal polarizations
per mode. We used the notation $\left\{ x:p\left(x\right)\right\} $
which refers to the set of $x$ for which $p\left(x\right)$ is true.
We may write (\ref{eq:SD}) as

\begin{equation}
\mathrm{DOS}\left(\omega\right)d\omega=\underset{\left\{ \left(m,n,l\right):f\left(\frac{2\pi m}{L},\frac{2\pi n}{L},\frac{2\pi l}{L},\omega<\omega^{\prime}\leq\omega+d\omega\right)=0\right\} }{\sum\frac{2}{L^{3}}\triangle m\triangle n\triangle l}\label{eq:SD2}
\end{equation}
since $m,n,$ and $l$ are all integers, that is $\triangle m=\triangle n=\triangle l=1.$
Replacing (\ref{eq:disp}) in (\ref{eq:SD2}) and converting the summation
to the integration (since L is large) gives
\begin{equation}
\mathrm{DOS}\left(\omega\right)d\omega=\frac{2}{\left(2\pi\right)^{3}}\underset{\left\{ \left(k_{x},k_{y},k_{z}\right):f\left(k_{x},k_{y},k_{z},,\omega<\omega^{\prime}\leq\omega+d\omega\right)=0\right\} }{\iiintop dV_{k}}\label{eq:DOS}
\end{equation}
where $dV_{k}=dk_{x}dk_{y}dk_{z}$ is the volume differential in k-
space. The integration in (\ref{eq:DOS}) is a volume (three-fold)
integration between the two 3D surfaces in k- space, satisfying dispersion
equation (\ref{eq:disp1}) at $\omega$ and $\omega+d\omega$, respectively.
These surfaces are also known as the iso-frequency surfaces. Solution
of equation (\ref{eq:DOS}) gives the DOS. Equation (\ref{eq:DOS})
holds regardless of the shape of the iso-frequency surfaces, and can
be used directly for simple surfaces. For example, since the iso-frequency
surfaces of a homogeneous isotropic space are spherical, using $k_{x}^{2}+k_{y}^{2}+k_{z}^{2}=k^{2}$
in (\ref{eq:DOS}) gives 

\begin{equation}
\mathrm{DOS}\left(\omega\right)d\omega=\frac{2}{\left(2\pi\right)^{3}}\intop_{\frac{\omega}{c}\sqrt{\varepsilon_{r}\mu_{r}}}^{\frac{\left(\omega+d\omega\right)}{c}\sqrt{\varepsilon_{r}\mu_{r}}}\intop_{0}^{\pi}\intop_{0}^{2\pi}k^{2}sin\theta d\varphi d\theta dk\label{eq:FS2}
\end{equation}
which after simple manipulation leads to the known equation $\mathrm{DOS}\left(\omega\right)=\frac{\omega^{2}\left(\mu_{r}\varepsilon_{r}\right)^{3/2}}{\pi^{2}c^{3}}.$

\begin{figure}
\centering{}\includegraphics[width=6.5cm]{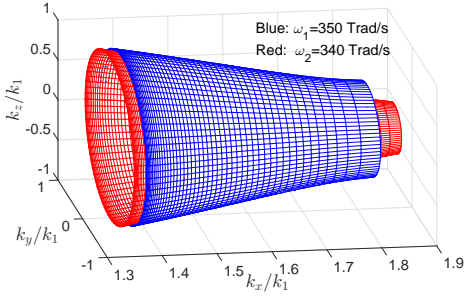}\caption{\label{fig:The-iso-frequency-surfaces}The iso-frequency surfaces
of a 1D PC (TE modes) at two different frequencies normalized to $k_{1}=\omega\sqrt{\mu_{0}\varepsilon_{0}\varepsilon_{1}}.$
Parameters of the 1D PC are $T_{1}=300\:nm,$ $T_{2}=700\:nm,$ $\varepsilon_{1}=2.1,$
$\varepsilon_{2}=11.9.$}
\end{figure}

We may prepare (\ref{eq:DOS}) further for an arbitrary iso-frequency
surface in k-space. This preparation becomes useful in finding $\mathrm{DOS_{Bloch}}\left(\omega\right)$
in photonic crystals and other non-trivial and anisotropic materials
such as hyperbolic metamaterials where iso-frequency surfaces'
shapes depend on the frequency. For example, Fig. \ref{fig:The-iso-frequency-surfaces}
shows the iso-frequency surfaces of a 1D PC at two nearby frequencies.

Each of the iso-frequency surfaces in (\ref{eq:DOS}) (and in Fig.
\ref{fig:The-iso-frequency-surfaces}) can be parametrized using two
independent parameters $\theta$ and $\varphi$ as

\begin{equation}
S\left(\omega\right):\:\left(k_{x}\left(\theta,\varphi\right),k_{y}\left(\theta,\varphi\right),k_{z}\left(\theta,\varphi\right)\right)\label{eq:surf}
\end{equation}
where $\left(k_{x},k_{y},k_{z}\right)$ represents a vector in k-space.
In general, the choice of parameters $\theta$ and $\varphi$ is arbitrary
and depends on the shape of the iso-frequency surface. However, it
is convenient to choose $\theta$ and $\varphi$ as spherical coordinate's
polar and azimuthal angles for 1D PCs. The surface differential for
such arbitrary surface is 

\begin{equation}
dA=\left\Vert t_{\theta}\times t_{\varphi}\right\Vert d\theta d\varphi\label{eq:A}
\end{equation}
where $t_{\theta}$, $t_{\varphi}$ are tangential vectors to the
surface as 

\begin{equation}
t_{\theta}=\left(\frac{\partial k_{x}\left(\theta,\varphi\right)}{\partial\theta},\frac{\partial k_{y}\left(\theta,\varphi\right)}{\partial\theta},\frac{\partial k_{z}\left(\theta,\varphi\right)}{\partial\theta}\right),
\end{equation}

\begin{equation}
t_{\varphi}=\left(\frac{\partial k_{x}\left(\theta,\varphi\right)}{\partial\varphi},\frac{\partial k_{y}\left(\theta,\varphi\right)}{\partial\varphi},\frac{\partial k_{z}\left(\theta,\varphi\right)}{\partial\varphi}\right),
\end{equation}
and $\left\Vert .\right\Vert $ and $\times$ are the vector $L^{2}$
norm the external vector product, respectively. Since $t_{\theta}\times t_{\varphi}$
is a vector normal to the surface, the volume differential between
the two iso-frequency surfaces at $\omega$ and $\omega+d\omega$
is 

\begin{equation}
dV=\left|\left(S\left(\omega+d\omega\right)-S\left(\omega\right)\right).\left(t_{\theta}\times t_{\varphi}\right)d\theta d\varphi\right|,
\end{equation}
which simplifies to 

\begin{equation}
dV=\left|\left(\frac{\partial k_{x}}{\partial\omega},\frac{\partial k_{y}}{\partial\omega},\frac{\partial k_{z}}{\partial\omega}\right).\left(t_{\theta}\times t_{\varphi}\right)d\theta d\varphi d\omega\right|.\label{eq:diff}
\end{equation}

Note the $\left|.\right|$ operator is necessary in the volume calculations
(since the volume between the two surfaces is independent of their
order). Replacing (\ref{eq:diff}) into (\ref{eq:DOS}) gives

\begin{equation}
\mathrm{DOS}\left(\omega\right)=\frac{2}{\left(2\pi\right)^{3}}\varoiintop_{S}\left|\left(\frac{\partial k_{x}}{\partial\omega},\frac{\partial k_{y}}{\partial\omega},\frac{\partial k_{z}}{\partial\omega}\right).\left(t_{\theta}\times t_{\varphi}\right)\right|d\theta d\varphi\label{eq:DOS_final}
\end{equation}
which can be used for any iso-frequency surface provided that it can
be parametrized. As a simple example, the spherical iso-frequency
surfaces of a non-magnetic ($\mu_{r}=1$) homogeneous isotropic medium
can be parametrized as 

\begin{equation}
S:\:\left(\frac{\omega\sqrt{\varepsilon_{r}}}{c}sin\theta cos\varphi,\frac{\omega\sqrt{\varepsilon_{r}}}{c}sin\theta sin\varphi,\frac{\omega\sqrt{\varepsilon_{r}}}{c}cos\theta\right)
\end{equation}
which easily leads to the expected result $\mathrm{DOS}\left(\omega\right)=\frac{\omega^{2}\varepsilon_{r}^{3/2}}{\pi^{2}c^{3}}.$

In a non-magnetic 1D PC, the dispersion relation is \cite{qi2017complex,joannopoulos1995photonic}

\begin{equation}
k_{x}=\frac{1}{T_{1}+T_{2}}cos^{-1}\left(cos\left(k_{x1}T_{1}\right)cos\left(k_{x2}T_{2}\right)-\quad\right.\label{eq:kx}
\end{equation}

\[
\qquad\qquad\left.0.5\left(\frac{p_{2}}{p_{1}}+\frac{p_{1}}{p_{2}}\right)sin\left(k_{x2}d_{2}\right)sin\left(k_{x1}d_{1}\right)\right),
\]
where 

\begin{equation}
k_{xi}=\sqrt{k_{i}^{2}-k_{y}^{2}-k_{z}^{2}};\;\quad p_{i}=\begin{cases}
\begin{array}{c}
\frac{k_{xi}}{\omega\mu_{0}}\\
\frac{\omega\varepsilon_{0}\varepsilon_{i}}{k_{xi}}
\end{array} & \begin{array}{c}
TE\\
TM
\end{array}\end{cases},
\end{equation}
and transverse electric (TE) and magnetic (TM) modes are defined with
respect to the y-z plane (interface plane).

The iso-frequency surface of the 1D PC can be parametrized as 

\begin{equation}
S\left(\theta,\varphi\right):\:k_{x}\left(\omega,\theta\right),\frac{\omega}{c}\sqrt{\varepsilon_{1}}cos\theta sin\varphi,\frac{\omega}{c}\sqrt{\varepsilon_{1}}cos\theta cos\varphi
\end{equation}
where $(\theta,\varphi)$ are defined inside the material with the
lower permittivity ($\varepsilon_1$) because the tangential wave-vector, $\sqrt{k_{y}^{2}+k_{z}^{2}}$,
inside the PC cannot exceed $\frac{\omega}{c}\sqrt{\varepsilon_{min}}.$
The tangential vectors to the iso-frequency surface are 

\begin{equation}
t_{\theta}:\:\left(\frac{dk_{x}\left(\omega,\theta\right)}{d\theta},-\frac{\omega}{c}\sqrt{\varepsilon_{1}}sin\theta sin\varphi,-\frac{\omega}{c}\sqrt{\varepsilon_{1}}sin\theta cos\varphi\right),\label{eq:tangential_PC_1}
\end{equation}

\begin{equation}
t_{\varphi}:\:\left(0,\frac{\omega}{c}\sqrt{\varepsilon_{1}}cos\theta cos\varphi,-\frac{\omega}{c}\sqrt{\varepsilon_{1}}cos\theta sin\varphi\right).\label{eq:tangential_PC_2}
\end{equation}

Using (\ref{eq:DOS_final}), (\ref{eq:tangential_PC_1}), and (\ref{eq:tangential_PC_2}),
it is straight forward to show the  density of Bloch states for TE and
TM modes are

{\scriptsize{}
\begin{equation}
\mathrm{DOS}_{\mathrm{Bloch}}^{i}\left(\omega\right)=\frac{\omega\varepsilon_{1}}{4c^{2}\pi^{2}}\intop_{0}^{\pi}\left|\left(cos\theta\frac{\partial k_{x}^{i}\left(\omega,\theta\right)}{\partial\theta}+\omega sin\theta\frac{\partial k_{x}^{i}\left(\omega,\theta\right)}{\partial\omega}\right)cos\theta\right|d\theta,\label{eq:DOS_PC}
\end{equation}
}where $i=\mathrm{TE}\:\mathrm{or}\:\mathrm{TM}$. Equation (\ref{eq:DOS_PC})
can be solved, as is, using commercially available solvers and no
further analytical expansion is necessary. However, special care should
be given to the calculation of $k_{x}$ using (\ref{eq:kx}) as it
passes through different Riemann sheets, as will be further discussed
later. 

\begin{figure}
\begin{centering}
\includegraphics[width=6cm]{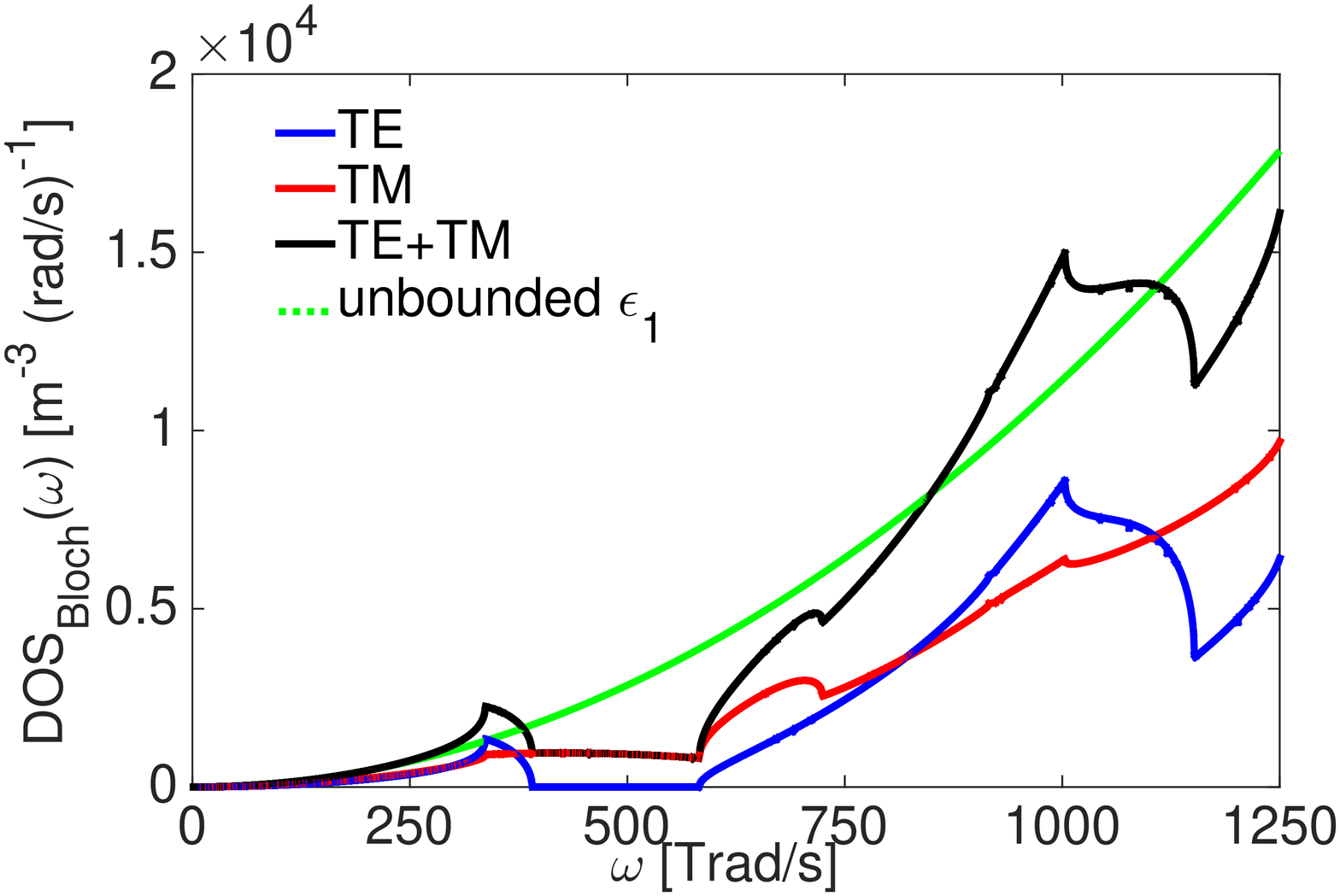}
\par\end{centering}
\caption{\label{fig: DOS-NOS} The Bloch density of states in the 1D PC. The
DOS of an isotropic homogenous $\varepsilon_{1}=2.1$ medium is also
shown as a reference.}
\end{figure}

As an example, Fig. \ref{fig: DOS-NOS} shows the  density of Bloch states inside a 1D
PC with parameters $\varepsilon_{1}=11.9$,
$\varepsilon_{2}=2.1$, $T_{1}=300\,\mathrm{nm}$, and $T_{2}=700\,\mathrm{nm}$. It also includes $\mathrm{DOS}$ of an isotropic homogeneous
material with $\varepsilon_{1}=2.1$. Note that Bloch wave's tangential
wave-vector inside a 1D PC is always limited by the material with
the lower permittivity (i.e. $\varepsilon_{1}$). The $\mathrm{DOS_{Bloch}}$
in Fig. \ref{fig: DOS-NOS} shows some dirivative discontinuities
which are at frequencies near the band edges of TM or TE modes. Some
of these local peaks exceed the DOS of an unbounded $\varepsilon_{1}$
region. 

\begin{figure}
\begin{centering}
\includegraphics[width=6.5cm]{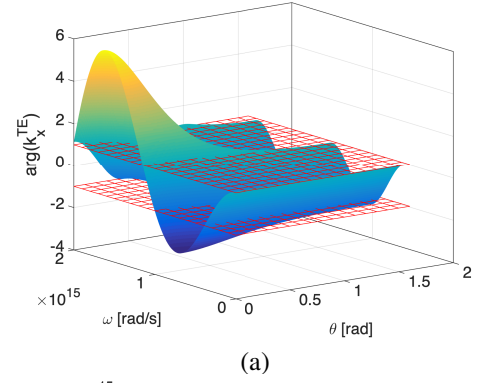}
\par\end{centering}
\begin{centering}
(a)
\par\end{centering}
\begin{centering}
\includegraphics[height=5cm]{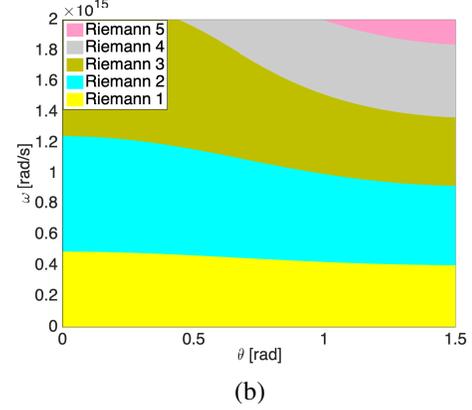}
\par\end{centering}
\begin{centering}
(b)
\par\end{centering}
\begin{centering}
\includegraphics[height=5cm]{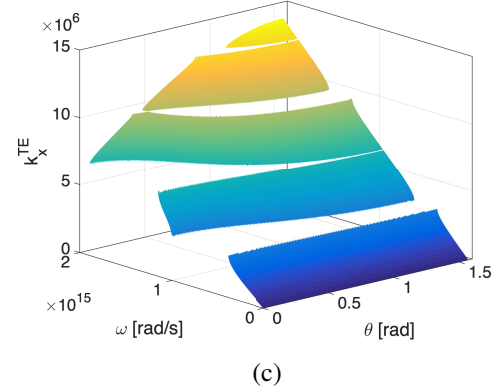}
\par\end{centering}
\begin{centering}
(c)
\par\end{centering}
\centering{}\caption{\label{fig: argument-of}a) argument of $cos^{-1}$ in \ref{eq:kx}
for TE modes. Flat surfaces (red) at $\pm1$ are added as a reference,
b) regions with different Riemann sheets, and c) $k_{x}^{TE}$ after
removing regions with imaginary $k_{x}^{TE}$ (band-gaps). The 1D
PC parameters are $d_{1}=300\:nm,$ $d_{2}=700\:nm,$ $\varepsilon_{1}=11.9,$
$\varepsilon_{2}=2.1.$}
\end{figure}

Note that the inverse cosine function in (\ref{eq:kx}) is a multi-valued
function with branch cuts on positive and negative sides of the real
axis ( $[-\infty,-1]$ and $[1,+\infty]$ segments). Every pass of
the inverse cosine argument across a branch cut requires considering
a different (appropriate) Riemann sheet. The argument of the $cos^{-1}$
function in (\ref{eq:kx}) for TE modes of the 1D PC example discussed
here is shown in Fig. (\ref{fig: argument-of}). The crossing points
of this function through branch cuts are where its $\frac{\partial}{\partial\omega}\left(.\right)=0.$
This leads to the Riemann sheet assignment shown in Fig. 5 (b) so
that 

{\small{}
\begin{equation}
cos^{-1}\left(x\right)=\begin{cases}
\begin{array}{c}
\left(n-1\right)\pi+\mathrm{P.V.}\\
n\pi-\mathrm{P.V.}
\end{array} & \begin{array}{c}
\mathrm{@Riemann}\;n,\quad\mathrm{for\:}\mathrm{odd\:n}\\
@\mathrm{Riemann}\;n,\quad\mathrm{for\:}\mathrm{even\:n}
\end{array}\end{cases}
\end{equation}
}where P.V. is the principal value of $cos^{-1}\left(x\right)$ as
$0\leq P.V.\leq\pi.$ The resulting $k_{x}^{TE}$ is also shown in
Fig. \ref{fig: argument-of}. 

\section*{Conclusion}

The density of Bloch states inside a 1D PC was formulated  based on
its dispersion relations for both TE and TM modes. The quantities
were calculated for a practical 1D PC in the visible range. 
\section*{Appendix}

The ratio of the radiation to absorption, $F\left(\omega,T\right),$
is considered to be a universal function for all solid matters based
on Kirchhoff's law (controversies regarding Kirchhoff's law are beyond
the purpose of this paper,  see \cite{robitaille2009kirchhoff,robitaille2003validity}
for details.) There are several methods to obtain $F\left(\omega,T\right),$
a.k.a. the black body radiation spectrum, inside a black body cavity.
They all lead to

\begin{equation}
F\left(\omega,T\right)=\rho\left(\omega\right)U\left(\omega,T\right)\label{eq:Kirchhoff}
\end{equation}
where $F\left(\omega,T\right)$ is considered to be a continuous function
of $\omega$ with the unit of joules per frequency per volume ($\frac{J}{m^{3}Hz}$),
and $U$ is the average total energy of the oscillators inside the
black body (or photons in the cavity). In the state of thermal equilibrium,
$U$ can be obtained either using the equipartition theorem in classical
statistical mechanics ($U=kT,$ where $k$ is Boltzmann's constant)
leading to Rayleigh-Jeans distribution, or using Plank's energy quantization
arguments leading to 
\begin{equation}
U=\hbar\omega / \left( exp(\frac{\hbar\omega}{kT})-1\right ).\label{eq:Boltzmann}
\end{equation}

One of the most intuitive methods to obtain  $\rho\left(\omega\right)$ in (\ref{eq:Kirchhoff}) (for isotropic homogeneous materials) is to count the supported electromagnetic
modes inside the resonator in the interval $\left[\omega,\omega+d\omega\right]$,
as we did in the main text. It can also be obtained from the emission
by classical resonators into an unbounded isotropic medium (modeling
a very large cavity) as follows \cite{milonni2013quantum}. Consider a particle with mass $m$ and charge $e$ acted upon by an
elastic restoring force $-m\omega_{0}^{2}z$ and an external electric
field of $E_{z}\left(t\right).$ For simplicity, assume the particle
only moves in one dimension ($z$). Newton's equation of motion for
such particle is 

\begin{equation}
\ddot{z}+\omega_{0}^{2}z=\frac{e}{m}E_{z}\left(t\right)+\frac{e}{m}E_{RR}\left(t\right)
\end{equation}
where $E_{RR}\left(t\right)$ is the reaction electric field originated
from the moving particle itself, and can be shown to be $E_{RR}\left(t\right)=\frac{2e}{3c^{3}}\dddot{z}\left(t\right).$
If the particle is inside a large rectangular cavity (with size L),
the energy absorption rate by the particle oscillating with frequency
$\omega_{0}$ from a cavity resonant mode at frequency $\omega$ is

\begin{equation}
\dot{W_{A}}=\frac{e^{2}}{2m}\frac{\gamma\omega^{4}E_{z}^{2}\left(\omega\right)}{\left(\omega^{2}-\omega_{0}^{2}\right)^{2}+\gamma^{2}\omega^{6}};\quad\gamma=\frac{2e^{2}}{3mc^{3}}\label{eq:absorption}
\end{equation}
where $E_{z}\left(\omega\right)$ is the $z-$directed electric field
associated with the electromagnetic mode, and $c$ is the speed of
($z-$polarized) light in the medium. If the radiation has a continuous
broadband spectrum, with the (z-directed electric field) energy density
of $F_{z}\left(\omega,T\right)d\omega=E_{z}^{2}\left(\omega\right)/\left(8\pi\right)$
in the interval $\left[\omega,\omega+d\omega\right]$, it can be shown
that \cite{milonni2013quantum}

\begin{equation}
\dot{W_{A}}=\frac{4\pi e^{2}}{m}\intop_{\omega}\frac{\gamma F_{z}\left(\omega,T\right)d\omega}{\left(1-\omega_{0}^{2}/\omega^{2}\right)^{2}+\gamma^{2}\omega^{2}}.\label{eq:abs_2}
\end{equation}

For the frequencies of interest in thermal emissions, $\gamma\omega\ll1$.
If $F\left(\omega,T\right)$ is not sharply peaked, we may simplify
(\ref{eq:abs_2}) as
\begin{equation}
\dot{W_{A}}=\frac{4\pi e^{2}\gamma F_{z}\left(\omega_{0},T\right)}{m}\intop_{\omega}\frac{d\omega}{4\omega_{0}^{-2}\left(\omega-\omega_{0}\right)^{2}+\gamma^{2}\omega_{o}^{2}}\label{eq:ans3}
\end{equation}

\[
\qquad=\frac{4\pi e^{2}\gamma F_{z}\left(\omega_{0},T\right)}{m}\frac{\omega_{0}^{2}}{4}\frac{2\pi}{\gamma\omega_{0}^{2}}.
\]

The energy emission rate of the oscillating charge is 

\begin{equation}
\dot{W_{E}}=\frac{2e^{2}\omega_{0}^{2}}{3mc^{3}}U\label{eq:emission-1}
\end{equation}
where $c$ is the speed of light in vacuum, and $U$ is the average
total oscillator energy. In the state of thermal equilibrium between
the radiational energy and the matter, (\ref{eq:emission-1}) and
(\ref{eq:ans3}) should be equal, leading to 

\begin{equation}
F\left(\omega_{0},T\right)=\frac{\omega_{0}^{2}}{\pi^{2}c^{3}}U,\label{eq:rad_}
\end{equation}
where $F_{x}\left(\omega_{0},T\right)=F_{y}\left(\omega_{0},T\right)=F_{z}\left(\omega_{0},T\right)=F\left(\omega_{0},T\right)/3$
is used. Similar to the resonator mode counting method, (\ref{eq:rad_})
gives 

\begin{equation}
\rho\left(\omega\right)=\frac{\omega^{2}}{\pi^{2}c^{3}}.\label{eq:RJ}
\end{equation}
Also, equations (\ref{eq:Kirchhoff}), (\ref{eq:Boltzmann}), and
(\ref{eq:RJ}) provide the well-known Plank's emission spectrum inside
a blackbody cavity filled with an isotropic homogeneous material. 

This method of obtaining $\rho\left(\omega\right)$ provides insight
into how radiators inside the matter (e.g. blackbody) couple to the
electromagnetic modes in common situations (leading to Plank's radiation
spectrum). Two critical assumptions which simplified (\ref{eq:abs_2})
to (\ref{eq:ans3}) are 1) radiation spectrum, $F_{z}\left(\omega,T\right),$
does not have any sharp peaks, and 2) the resonators' coupling to
the radiation is sharply peaked around their natural frequency, $\omega_{0}.$
This assumption also implicitly requires linearity of the material.
Only with these assumptions we obtained the same $\rho\left(\omega\right)$
as using the other, more fundamental, methods such as BE distribution
of photons in the cavity in thermal equilibrium. It appears that (\ref{eq:abs_2})
is a good starting point to study the thermal emission from non-linear
(or any other uncommon) material.

\bibliographystyle{IEEEtran}


\end{document}